\documentclass[pra,aps, twocolumn,showpacs]{revtex4}

\usepackage{exscale}
\usepackage{xspace}
\usepackage{citesort}

\usepackage{amsfonts}
\usepackage{amsmath}
\usepackage{amssymb}
\usepackage[dvips]{graphicx}
\usepackage{subfigure}

\newcommand{\ket}[1]{\left| #1 \right\rangle}
\newcommand{\bra}[1]{\left\langle #1 \right|}

\newcommand{\ot}{{\,\otimes\,}}
\graphicspath{%
    {converted_graphics/}
    {/}
}
\begin{document}

\title{ Quantum Zeno effect on Quantum Discord }

\author{ A. Thilagam} 
\address{Information Science, Engineering and Environment, 
Mawson Institute,
University of South Australia, Australia
 5095} 
\date{\today}
\begin{abstract}
We examine the quantum  Zeno effect on the dynamics of quantum discord in
two initially entangled qubits  which are  subjected 
to frequent measurements via decoherent  
coupling with independent reservoirs. 
The links between  characteristic parameters 
such as system bias,  measurement time duration, 
 strength of initial entanglement between the 
two qubit systems and the dynamics of quantum  discord are  examined
for two initial state configurations. At weak or
unsharp measurements, the quantum discord, which is an intrinsically 
distinct entity from concurrence, serves as a reliable indicator 
of the crossover point in Zeno to anti-Zeno transitions. 
However at highly precise  quantum measurements, 
the monitoring device interferes significantly with the 
evolution dynamics of the monitored system, and 
the quantum discord yields indeterminate values in a reference
frame where the observer is not an active  constituent of 
the subsystems.
 
\end{abstract}
\maketitle

\section{Introduction}\label{c1a}

Recently, studies of separable and therefore non entangled states 
containing other kinds  of non classical correlations has attracted increased
attention. One such correlation measure, the quantum discord \cite{zu,ve1,ve2},
based on the difference between quantum and classical information theories, 
 incorporates more generalized correlations not seen in 
other non-classical correlations such as  entanglement. 
 In particular quantum states with zero entanglement 
properties are seen to possess quantum discord  and 
classical-quantum states which are  necessarily separable have zero quantum discord. 
Two positive discord states
can be mixed to obtain a zero-discord classical state, and two
zero-discord classical states in  orthogonal directions
can be merged to form a non-zero discord state \cite{matt}.
Moreover the quantum discord is not restricted by the monogamy rule \cite{woot}
which is obeyed by the concurrence measure during  entanglement sharing.
Such  intriguing features of quantum discord  has opened 
up avenues for variety of attributes and applications  in non-markovian 
open quantum systems \cite{tern,ferr,fan,maz,pii}, spin array systems \cite{cil}
detection of  quantum phase transitions \cite{wer}, 
 quantum information 
processing \cite{datta} and quantum communication \cite{pia}. 

The quantum Zeno effect (QZE) describes the retarded time evolution 
of a quantum state subjected to frequent measurements\cite{Misra,It,FacJ}. 
In the limiting case of continuous measurement, the time evolution 
of the state comes to a standstill. The opposite
effect which leads to enhancement in time evolution is known 
as anti-Zeno effect (AZE)  and has been observed 
to be much more ubiquitous than Zeno effect \cite{ob}. 
In unstable systems, the occurrence of 
both QZE and AZE effects depends on critical parameters 
like measurement frequencies and environmental noise \cite{env}.
Quantum systems  exhibiting  both effects
 include the nanomechanical oscillator \cite{nosci}, 
two-state system coupling to a
spin chain environment in transverse magnetic fields \cite{wangS},
the non equilibrium steady state spin-fermion model 
a variant of the Kondo model \cite{Segal}, 
damped quantum harmonic oscillator \cite{env}, disordered
spin systems \cite{japko}
and trapped  atomic systems \cite{rai}. The
 nanomechanical oscillator system, in  particular is of increased
 interest as it provides
an ideal medium for testing quantum effects on a macroscopic scale.

An inherent feature in determining quantum discord involves  the one-sided 
projective measurements on a selected subsystem of the composite
quantum state. As is well known, this introduces various 
counter-intuitive
features linked with the measurement process itself with associated 
controversies linked with the collapse
of the wave function of the measured system.
For instance the term 
``subjective reality" was introduced by Wiseman \cite{wise} 
to describe the dependence of quantum trajectories
on the observer's measurement frame, hence there are several ways that 
quantum systems which are monitored  can be interpreted. 
 A well-known approach to the widely used collapse postulate
involves its replacement by the decoherence process
subjected by a detector on the system under study \cite{zuro}
An alternative scheme involves the idea of  quantum Zeno 
subspaces \cite{FacJ} which provides  a convenient platform
for interpreting the Zeno effect.
In this regard, we note that the active presence of the
measuring device  is not a requirement for  quantum 
Zeno effects to be seen. This is due to the fact that 
the Zeno effect is linked to the 
evolution of the non-Hermitian Schr\"odinger equation associated
with any irreversible mechanism, with the act of measurement
being a well known one.
 
 For low precision or unsharp measurements,
the device $D$ introduces  minimal disturbance on the measured system, 
$S$ with state $u|S_u\rangle + v|S_v\rangle$. 
The state of the measuring device can be $|D_u\rangle$ or 
$D_v\rangle$ after the measurement, and  is different from
its state before measurement, $|D_i\rangle$. The composite
system $S \otimes D$ proceeds in an  approximately unitary 
fashion as $U |S_u\rangle |D_i\rangle = |S_u\rangle |D_u\rangle$,
$U |S_u\rangle |D_i\rangle = |S_u\rangle |D_u\rangle$.
In the case of ideal measurements, the resulting state of the
system after measurement  generally belongs to the set
of the orthonormal basis of the quantum system.
Thus for weak or unsharp measurements,
the non-Hermitian term can be ignored and simplified
approaches such as that based on the Kofman and Kurizky's formalism \cite{ob} 
can be employed to analyze the effect of measurements.

For highly precise measurements, any analysis of the quantum evolution 
becomes complicated due to the  influence of the  non-Hermitian term, which can
be interfere strongly with the dynamics of the
measured system. Accordingly, we provide an 
 analysis of the evolution of a measured system
involving a non-Hermitian term which appears due to 
highly precise measurements or a strong monitoring device here.
This is performed by applying the results of the non-Hermitian Hamiltonian of a two-level 
system originally solved in the context of the link between a decay term 
and Berry's phases by Garrison and Wright \cite{gar} to our measurement 
model. We note that a analogous decay  term is explicitly linked with the
precision of quantum measurements, a higher  measurement precision results
in a larger magnitude of this decay term. Some ideas introduced in this
work may thus be extended to study the links
 between Berry phases and the quantum measurement problem.

For ideal or weak measurements, the  
Von Neumann projection operator ${\cal P}$   \cite{Misra,FacJ} is convenient 
to formulate measurement procedures in Hilbert space ${\cal H}$ of a quantum system, $S$.
The initial density matrix $\rho_0$ of system $S$  
is constrained within  ${\cal H}_{\cal P}$ as 
$\rho_0 = {\cal P} \rho_0 {\cal P} ,\; \; \mathrm{Tr} [ \rho_0 {\cal P} ] = 1$.
In the absence of any measurement, the state evolves as
$\rho (t) = U(t) \rho_0 U^\dagger (t)$ where 
$U(t)=\exp(-iH^\star t)$, and  $H^\star $ 
is  a time-independent Hamiltonian.
The probability that the
system remains within  ${\cal H}_{\cal P}$ is given by 
$P(t) = \mathrm{Tr} \left( U(t) \rho_0 U^\dagger(t) {\cal P}\right)$.
In the event of measurement at time $\tau$, density matrix $\rho(\tau)$
transforms as  $ \rho(\tau) = \frac{1}{P(\tau)}\;{\cal P} U(\tau) \rho_0 U^\dagger(\tau)
{\cal P}$. The survival probability in ${\cal H}_{\cal P}$ is given by
$P(\tau) = \mathrm{Tr} \left(V(\tau) \rho_0 V^\dagger(\tau)
\right)$ where $V(\tau) \equiv {\cal P} U(\tau){\cal P}$. For 
measurements  taken at time intervals $\tau=t/N$,
the survival probability is given by 
\begin{eqnarray}
\label{eq:survorig}
P^{(N)}(t) = \mathrm{Tr} \left( V_N(t) \rho_0 V_N^\dagger(t)\right),\\
\nonumber 
 V_N(t) = \left[ V\left(\frac{t}{N}\right)\right]^N
\end{eqnarray}
At very large  $N$, no   transitions  allowed outside ${\cal H}_{\cal P}$ 
occur and $ P^{(N)}(t) \rightarrow 1$, the culmination of the mathematical
formulation of the Zeno effect. 
Eqs. (\ref{eq:survorig}) embodies the intriguing effect of a measurement process,
where a system monitored to determine whether it 
remains in a particular state persists to remain in that state.
 This idea has been examined via the adiabatic theorem 
 \cite{FacJ} in which different outcomes 
are eliminated and the system evolves
as a group of exclusive quantum Zeno subspaces within the 
total Hilbert space. 
The measurement procedure therefore  has a decomposing
effect on the total Hilbert space which is partitioned
into orthogonal quantum Zeno subspaces \cite{FacJ}.
The initial state remains in 
a particular invariant subspace, and 
its survival probability remains unchanged over a period of time.

The effect of measurement on the dynamics of quantum discord can be examined
in one of several ways. 
An obvious one  involves examining the role of Zeno effect associated with 
measurements introduced in one subsystem in order  to  obtain the conditional
entropy, and enabling determination of  the classical correlation measure based on optimal 
measurements. This procedure forms 
the basis of determining the quantum discord,  as shown in earlier mathematical
 formulations \cite{zu,ve1,ve2}. In order to evaluate the quantum discord,
a set of positive-operator-valued measurements (POVM) need to be performed
in a neighboring partition. 
Does the measurement process itself  induce a distinct category of 
quantum discord? 
How exactly can such optimal measurements be performed
without incurring the quantum Zeno effect?
What are the key attributes of an optimal measurement and the possible
role played by the Zeno effect in POVM?  
In this regard, the consideration of distinct  measurement
 techniques in separate sub-systems will introduce greater depth to the analysis 
of the quantum discord present in the global system. 
This includes the effects due to the asymmetry of measurement procedures.
However such detailed investigations is not an easy task, as 
the difficulty in determining the quantum discord even for simpler
systems is well known. So far 
analytical form has been derived only under restricted
conditions \cite{luo,sara,ali,ali2}.

For the sake of obtaining analytical expressions, it is generally 
assumed that the measurement time duration 
or frequency of measurements is the same for  subsystems
not in contact with any reservoir system. We continue to assume
 this model for simplicity in analytical
treatment, however  we opt to examine the effect of measurement from a different 
perspective. This involves examining the influence of 
the Zeno-like  effect associated with acts of continuous measurements
 by the environment that  is in contact with 
the qubit subsystem \cite{koro}. 
The well-known model of the
 solid-state qubit interacting with a reservoir system
presents a convenient platform for examining the complicated link
between quantum Zeno effect, quantum discord and 
the dynamics of Zeno subspaces. The reservoir may be viewed
as providing the  ``back-action" needed for the
dynamical collapse of the  wave-function collapse.

 In order to keep the problem tractable,
we consider in the first  instance, the well-known model of 
a pair of  initially entangled
spin-boson system with independent harmonic  reservoirs found
 useful in quantifying salient aspects of dissipative dynamics
 of many quantum systems \cite{Leg,Weiss}.
Factors such as  spectral density, bias and temperature
are considered to play important roles in the overall dynamics of the
qubit-reservoir system. We follow Prezhdo's approach involving the 
quantum control of chemical reactivity by a solvent acting as the environment
\cite{prez},  the anti-Zeno mechanism therefore occurs  by loss of electronic coherence 
in some chemical systems.
 The interplay of various quantum interactions (non-local and local)
between the environment and the qubit system results in 
the reservoir acting as continuous detector.

Our paper is organized as follows. In Section \ref{meas} we provide 
a brief review the concept
of quantum discord and highlight the role of measurements in its formulation.
In Section \ref{c1b}, we describe salient features of Zeno dynamics of   
the spin-boson system using Kofman and Kurizky's formalism \cite{ob} which yields the
effective decay of a quantum system under ideal measurements.
In Section \ref{dyn} we investigate the influence of quantum Zeno effect
on the dynamics of the quantum discord for X-type qubit states
 with two initial state configurations.
We present our main results and make comparisons between the quantum discord and 
the concurrence measure. 
In Section \ref{exp}, we analyze 
the non-Hermitian dynamics resulting from highly precise measurements
on a two-level quantum system, and highlight the appearance of exceptional points.
A brief discussion and conclusions are then presented in Section \ref{con}.

\section{Measurements and Quantum discord}\label{meas}

Following the formulation of quantum discord in 
Refs.\cite{zu, ve1, ve2}, we express
the quantum mutual information of a composite state $\rho$ of
 two subsystems $A$ and $B$  as
$\mathcal{I}(\rho) = S(\rho_A) + S(\rho_B) - S(\rho)$
for a density operator in $\mathcal{H}_A
\ot \mathcal{H}_B$. $\rho_{A}$ and $(\rho_B)$ are 
reduced density matrices and $S(\rho_i)$ (i=A,B) denotes 
the well known  von Neumann entropy of the density operator $\rho_i$.
$S(\sigma)= - {\rm tr}(\rho \log\rho)$ stands
The mutual information can also be written in terms of 
quantum conditional entropy $S(\rho|\rho_A)= S(\rho) - S(\rho_A)$
as $\mathcal{I}(\rho) = S(\rho_B)  - S(\rho|\rho_A)$.

 The quantum Zeno effect appears as a result of the
 measurement process intrinsic 
in the definition of the 
conditional entropy.
A series of one-dimensional orthogonal projectors $\{\Pi_k\}$ induced
in $\mathcal{H}_A$ leads to different outcomes of
the measurement. We are presented with the post  measurement conditional state \cite{bylic}
$\rho_{B|k} = \frac{1}{p_k} (\Pi_k \ot \mathbb{I}_B)\rho (\Pi_k \ot
    \mathbb{I}_B)$
where the probability $p_k = {\rm tr}[\rho_{B|k} (\Pi_k\ot \mathbb{I}_B)]$
and  $\{\Pi_k\}$ denote the one-dimensional projector indexed by the 
outcome $k$. A conditional
entropy of the subsystem $B$ can be attached to $\rho_{B|k}$ based 
on the cumulative effect of the mutually exclusive measurements 
on $A$ as $S(\rho|\{\Pi_k\}) = \sum_k p_k S(\rho_{B|k})$.
The measurement induced mutual information is therefore
$ \mathcal{I}(\rho|\{\Pi_k\}) = S(\rho_B) - S(\rho|\{\Pi_k\})$
 while the  classical correlation measure based on optimal 
measurements is obtained as \cite{zu,ve1,ve2}
$ \mathcal{C}_{A}(\rho) = \sup_{\{\Pi_k\}} \mathcal{I}(\rho|\{\Pi_k\})$.
The difference in $\mathcal{I}(\rho)$ and
$\mathcal{C}_A(\rho)$ yields the non symmetric term known as quantum discord, $  \mathcal{D}_{A}(\rho) =  
\mathcal{I}(\rho) - \mathcal{C}_A(\rho)$. A  discord
$\mathcal{D}_{B}(\rho)$ corresponding to measurements made on $B$ can likewise
be obtained and need not be the same as $\mathcal{D}_{A}(\rho)$. 
As to be expected, the quantum discord  is not symmetric with respect to $A$ and $B$,
 particularly if attributes such as the measurement duration employed in either subsystems
differ. 

 \section{Zeno dynamics of   the spin-boson system}\label{c1b}

In order to examine the dynamics of   the spin-boson system, we utilize the density matrix 
associated with the Liouville equation $\frac{\partial \rho}
 {\partial t} = - i [\widehat  H_{\rm T},\rho(t)]$,
where the total Hamiltonian $\widehat  H_{\rm T} =\widehat  H_{\rm qb}+
\widehat H_{\rm os} + \widehat  H_{\rm qb-os}$ and $\widehat  H_{\rm qb}$
of the two-level qubit assumes the form $\widehat  H_{\rm qb}
= \hbar \left(\frac{\Delta \Omega}{2}\, \sigma_{z} +\Delta \sigma_{x}\right)$.
The Pauli matrices are expressed in terms of the two possible
states $(\ket{0},\ket{1})$, $\sigma_{x} = \ket{0} \bra{1}
+ \ket{1} \bra{0}$ and $\sigma_{z} = \ket{1} \bra{1}
- \ket{0} \bra{0}$. $\Delta \Omega$ is the biasing energy
while  $\Delta$  is the tunneling  amplitude. 

We consider that the  two  uncoupled qubits  are  coupled to 
independent reservoirs  of harmonic oscillators,
$\widehat  H_{\rm os} = \sum_{\bf q} \hbar \omega_{\bf q}\,
b_{\bf q}^{\dagger}\,b_{\bf q}$. $b_{\bf q}^{\dagger}\,$ and $b_{\bf q}\,$
are the respective  creation and  annihilation 
operators of the quantum oscillator with wave vector ${\bf q}$. 
The qubit-oscillator interaction Hamiltonian is 
linear in terms of oscillator  creation and annihilation operators
$\widehat  H_{\rm qb-os} = 
\sum_{{\bf q}}   \lambda_{_{\bf q}}\, \left ( b_{\bf q}^\dagger +
b_{\bf q}\right ) \sigma_{z}$.
The term $\lambda_{_{\bf q}}$ denotes the coupling between the qubit and the
environment and is characterized by the spectral density function,
$J(\omega)=\sum_{\bf q}\lambda_{_{\bf q}}^2\delta(\omega-\omega_{\bf q})$,  which 
we assume to be of the ohmic 
form $J(\omega)= 2 \pi \eta \omega e^{-\frac{\omega}{\omega_c}}$.
 $\eta$ is the dimensionless reservoir coupling function, and  $\omega_c$ is 
the reservoir cutoff frequency. We 
consider the measuring device to be an active constituent of  
the total Hamiltonian $\widehat  H_{\rm T} =\widehat  H_{\rm qb}+
\widehat H_{\rm os} + \widehat  H_{\rm qb-os}$. The
 reservoir assumes  the role of the 
 measuring device, by inducing 
 a projection operation that disrupts the normal evolution of
Hamiltonian $\widehat  H_{\rm T}$. 
The reservoir here  serves the same role as the solvent in 
Prezhdo's work on the 
quantum control of chemical reactivity \cite{prez}.

Each qubit decays  
to oscillator states in the reservoir when measurements are made,
making  a transition from  its excited state
$\ket{1}_{\mathrm{q}}$ to ground state $\ket{0}_{\mathrm{q}}$.
 We consider an initial state of the qubit with  its corresponding 
reservoir in the vacuum state, existing in  equilibrium at temperature $T=0$K
$
|\phi _{i}\rangle =|1\rangle _{\mathrm{q}}\otimes
\prod_{k=1}^{N'}|0_{k}\rangle _{\mathrm{r}}=
|1\rangle _{\mathrm{q}}\otimes
\ket{{\bf 0}}_{\mathrm{r}}$ 
where $\ket{{\bf 0}}_{\mathrm{r}}$ implies that all $N'$ wavevector modes 
of the reservoir are unoccupied in the initial state.
$|\phi _{i}\rangle$ then   undergoes  the 
following mode of decay 
\begin{equation}
|\phi _{i}\rangle \longrightarrow
 u(t) \; \ket{1}_{\mathrm{q}}
\ket{{\bf 0}}_{\mathrm{r}} + v(t) \; \ket{0}_{\mathrm{q}}
\ket{{\bf 1}}_{\mathrm{r}} ,  
\label{fstate}
\end{equation}
In order to keep the problem tractable we
consider that $\ket{{\bf 1}}_{\mathrm{r}}$ denotes
a  collective state of  the reservoir,  $
|{\bf 1}\rangle _{\mathrm{r}}=\frac{1}{v(t)}
\sum_{n} \lambda _{\{n\}}(t)|\{n\}\rangle$ 
where $\{n\}$ denotes an occupation scheme in which there are
$n_i$ oscillators with wavevector $k=i$ in the reservoir
and we define the state $|\{n\}\rangle$ as $
|\{n\}\rangle =|n_0,n_1,n_2...n_i..n_{N'}\rangle$.

For  ideal measurements, 
the functions $u(t)$ and $v(t)$ in Eq.(\ref{fstate}) satisfy 
the relation $u(t)^2 + v(t)^2=1$, and can be considered
to be approximately  satisfied for
unsharp or weak measurements which introduce minimal disturbance to the
system being monitored. The square of the 
function $u(t)$  yields the survival probability
 associated with  $N$ measurements performed at
regular intervals $\tau$, $P(t)= u(t)^2 = \exp(-N \Delta^2 \tau^2/4)$ where $t=N \tau$.
In the extreme limit $\tau \rightarrow 0, \; u(t) \rightarrow 1$ and the decay into
phonon states is totally inhibited. For small $\tau, N$ values 
and a weak qubit-reservoir coupling, we assume that 
the state of the collective reservoir  at time $t= \tau$  is 
 equivalent to that at   $t= N \tau$. The second order processes giving rise to
exchanges between oscillators and  hence changes in  the ensemble   configuration
of oscillators  in can be considered  minimal and   neglected at small $t$. 
At very  short times,  the  effective relaxation rate for the 
two-level qubit is given by $\gamma(\tau)=(\Delta/2)^2\tau$ so that 
$u(t)^2 = \exp(-\gamma(\tau) \tau)$. 
The decay of  a quantum state interacting with a reservoir
is almost zero at the beginning of the decay process, a typical 
behaviour in quantum Zeno effect. At  intermediate  measurement time intervals, the
decay of  quantum state  may be accelerated as is the case in anti-Zeno effects.

It is to be noted that Eq.(\ref{fstate}) does not provide a 
dynamic description of the measurement process such
as the  evolution of the system during or after $N$ measurements.
Eq.(\ref{fstate}) stems from a 
  probabilistic interpretation of quantum measurements and
 predicts the two possible outcomes,  consistent
with Born's rule linked with the probabilistic nature of the
projection postulate.  The survival probability given by $|u(t)|^2$
is  the cumulative outcome of several successive measurements.
As is well known,  inconsistencies  still
remain in problems associated with  quantum measurements.
The unitary and reversible features of the Schr\"odinger
equation and  the non-unitary elements inherent in the projection
postulate are clearly incompatible. However both these core
 processes need to be unified in order  to examine the influence of a continuous monitoring  
on the unmeasured evolution of a quantum system, which is a challenging
task.

We evaluate the effective decay rate of the spin-boson model 
at small values of $\tau$,
using Kofman and Kurizky's formalism which is 
based on the  convolution of two functions \cite{ob}
\begin{equation}
\gamma(\tau)=2 \left(\frac{\Delta}{2}\right)^2 \int_0^{\infty} 
d\omega K(\omega) F_{\tau}(\omega-{\Delta \Omega}),
\label{eq:overlap}
\end{equation}
The function $F_{\tau}(\omega-{\Delta \Omega})
=\frac{\tau}{2\pi} {\rm sinc}^2\left[ \frac{(\omega- {\Delta \Omega})\tau}{2}\right]$ 
and is associated with measurements at intervals of $\tau$.
The reservoir coupling function $K(\omega)$ is evaluated using 
$K(\omega) = \int_0^{\infty} \; e^{i \omega t}
\cos[\Delta \Omega + G_1(t)] e^{-G_2(t)} d t $
where $G_1(t) = \int_0^{\infty}d\omega \frac{J(\omega)}{\omega^2}\sin\omega t$
and $ G_2(t) = \int_0^{\infty}d\omega \frac{J(\omega)}{\omega^2} 
\coth[\frac{\beta \omega}{2}] (1-\cos\omega t)$, 
where $\beta = \frac{1}{k_B T}$ and $T$ is the lattice temperature.
Explicit expressions for $G_1(t)$
and $G_2(t)$ in Refs.\cite{Leg,Weiss} for an ohmic $J(\omega)$
show the strong dependence of  $K(\omega)$ on  the reservoir coupling function $\eta$ and
the exponential cutoff frequency $\omega_c$.
The occurrence of QZE or AZE is determined by 
changes in the overlap between  functions $F_{\tau}(\omega)$ and $K(\omega)$ as
$\tau$ is varied.  QZE (AZE) occurs when the overlap of functions decreases (increases) 
with decrease in $\tau$. The crossover from QZE to AZE 
is most pronounced when ${\tau}$ is increased
in systems with weak spin-boson coupling \cite{thila}, and also
when bias $\Delta \Omega$ is increased as well.

\subsection{Approximate relations of the Zeno-anti Zeno crossover point}\label{ca}

To obtain approximate analytical relations, we employ the 
 an effective decay rate applicable at short times \cite{Segal},
$ \gamma(\tau)=  \frac{\Delta^2} {2\tau} \Re \int_0^{\tau}dt_1 \int_0^{t_1}K(t')dt'$
where $K(t)$  is the fourier transform of the 
coupling function $K(\omega)$ defined below Eq.(\ref{eq:overlap}).
Using explicit expressions for $G_1(t)$
and $G_2(t)$ given in Refs.\cite{Leg,Weiss} for an ohmic $J(\omega)$, 
we obtain $\gamma(\tau)$ as follows
\begin{eqnarray}
\label{eq:toy2}
\gamma(\tau)= \frac{\Delta^2}{2 } {(\frac{\pi}{\beta})}^{2 \eta}
  \int_0^{\tau} && dt
\cos[\Delta \Omega + 2 \eta \tan^{-1}\omega_c t] \\ \nonumber
&& \times \frac{t^{2 \eta}}{(1+ (\omega_c t)^2)^{\eta}}
{({\rm csch} \frac{\pi t}{\beta})}^{2 \eta}
\end{eqnarray}
where $\beta = \frac{1}{k_B T}$,  $T$ is the lattice temperature
and ${\rm csch(x)}$ is the hyperbolic cosecant function.
Using  $\Delta \Omega=0$,  $T=0$K, $\Delta^2 =2$ and $\omega_c=1$,
we  obtain  simple expressions for 
$\gamma(\tau)$ and $\frac{\partial \gamma(\tau)}{\partial \tau}$
\begin{eqnarray}
\nonumber
\gamma(\tau)= && \frac{(1+ \tau)^{-\eta}}{(2 \eta-1)}
\left (\sin[ 2 \eta \tan^{-1} \tau]-
 \tau \cos[ 2 \eta \tan^{-1} \tau] \right ) 
\\ \label{eq:toy3} \\ \label{eq:toy4} 
\frac{\partial \gamma(\tau)}{\partial \tau} && =
(1+ \tau)^{-\eta}  \cos[ 2 \eta \tan^{-1} \tau]
\end{eqnarray}
At very short time intervals $\omega_c \tau < 1$, $\gamma(\tau) \approx \tau$
 whereas at very large  times $\omega_c \tau \rightarrow \infty$ and for $\eta \neq  1/2$, 
$\gamma(\tau) \approx \frac{\cos \pi \eta}{2 \eta-1}\; \tau^{1-2 \eta}$. 
 At $\eta = 1/2$, $\frac{\cos \pi \eta}{1-2 \eta} \rightarrow \frac{\pi}{2}$ and
we get a rate which is independent of the measuring device, 
$\gamma(\tau) = \frac{\pi \Delta^2}{4 \omega_c}$. 
 
At the point of Zeno-anti-Zeno transition, 
$\frac{\partial \gamma(\tau)}{\partial \tau} = 0$, and using Eq.(\ref{eq:toy3})
we obtain an explicit  expression for 
the measurement interval $\tau_{_{T}}$ at which Zeno
to anti-Zeno transition occurs ($\eta \neq  1/2$)
\begin{equation}
\tau_{_{T}}=\tan \frac {\pi}{4 \eta}
\label{eq:rateT}
\end{equation}
At non-zero values of $\Delta \Omega$ where 
the spin-boson system exist under biased conditions,  
the measurement interval $\tau_{_{T}}$ at which Zeno
to anti-Zeno transition occurs is  modified to
\begin{equation}
\label{eq:rateTB} 
\tau_{_{T}}=\tan \left[ \frac{1}{2 \eta} \left(\frac{\pi}{2}- \mu \Delta \Omega \right)\right ]
\end{equation}
where the factor $2< \mu < 3$ and depends on the bias   $\Delta \Omega$.
Eq.(\ref{eq:rateTB}) is consistent with the fact that 
an increase in   biasing energy
$\Delta \Omega$ increases the probability of Zeno-anti-Zeno transition.

It is important to note that the defination of the Zeno-anti-Zeno transition 
is based on the properties of the decay rate, $\gamma(\tau)$ and not
a fixed natural rate. $\tau$ can be viewed as the duration of one of many other  pulses,
and therefore specific to the local dynamics of the quantum system being monitored.
We point out the  difference in settings  between the current work 
and an  earlier work \cite{thila} in which the reservoir
constitutes a part of the dynamical system that is monitored by the measuring device.
In Ref. \cite{thila},  Zeno to anti-
Zeno features were  revealed even with the first (of many) measurement by a distant observer 
 due to the continuous measurement effect by the reservoir of oscillators.

\section{\label{dyn} Dynamics of Quantum discord for X-type qubit states }

In order to examine the  joint evolution of a  pair of two-level qubit systems 
in uncorrelated reservoirs, we consider the following  Bell-like initial   state
\begin{eqnarray}
\ket{\Phi}_0 &=& \left[ a\ket{0}_{\mathrm{q1}}\ket{0}_{\mathrm{q2}}
+ b\ket{1}_{\mathrm{q1}}\ket{1}_{\mathrm{q2}} \right]
 \ket{0}_{\mathrm{r1}} \ket{0}_{\mathrm{r2}},
 \label{fstate2}
\end{eqnarray}
where $i$=$1,2$ denote the two qubit-reservoir systems associated
function $u_i(t)$ in  Eq.(\ref{fstate}). 
 $a,b$ are real coefficients and satisfy, $a^2+b^2=1$.
Using  Eq.(\ref{fstate})   and tracing  out the reservoir states we obtain
a  time-dependent qubit-qubit reduced density matrix 
in the basis $(\ket{0 \;0},\ket{0 \;1}\ket{1 \;0}\ket{1 \;1})$ which
evolves with time duration $\tau$ as
\begin{eqnarray}
\label{matrix1}
\rho_{_{\mathrm{q1,q2}}}(t)=
\left(
\begin{array}{cccc}
 f_1& 0 & 0 &f_5 \\
  0 &  f_2 & 0 & 0 \\
  0 & 0 & f_3 & 0 \\
  f_5& 0 & 0 & f_4\\
\end{array}
\right).
\end{eqnarray} 
where $f_1=a^2+b^2v_1(\tau)^2 v_2(\tau)^2$, $f_5=a b  u_1(\tau)\; u_2(\tau)$,
$f_2=b^2 v_1(\tau)^2 u_2(\tau)^2$, $f_3= b^2 u_1(\tau)^2  v_2(\tau)^2$, 
$f_4= b^2 u_1(\tau)^2 u_2(\tau)^2$.
We assume that the usual unit trace
and positivity conditions of the density operator
 $\rho_{_{\mathrm{q1,q2}}}$ are satisfied,
however  these may not constitute strict requirement for the determination
of the quantum discord.
The  reservoir-reservoir  reduced density matrix 
$\rho_{_{\mathrm{r1,r2}}}$ is similarly  obtained by 
by tracing out qubit states. Each  non-zero matrix term of 
$\rho_{_{\mathrm{r1,r2}}}$ is easily obtained from the corresponding 
term $ \rho_{_{\mathrm{q1,q2}}}(t)$  by swapping 
 $u_i \leftrightarrow v_i$. Both  matrices possess the well-known 
 $X$-state structure which preserve its form
during evolution. The  well known Wootters
 concurrence \cite{woot} for
the  density matrix in  Eq.(\ref{matrix1}) is
$\mathcal{C}_{_{\mathrm{q1,q2}}}(\tau)= 2 b e^{-{\frac{1}{2}}(\gamma_1+\gamma_2)\tau}
 \times \left[
a-b (1-e^{-\gamma_1 \tau})^{\frac{1}{2}}(1-e^{-\gamma_2 \tau})^{\frac{1}{2}} \right]$ and
$\mathcal{C}_{_{\mathrm{r1,r2}}}(\tau)=2 b(1-e^{-\gamma_1 \tau})^{\frac{1}{2}}
(1-e^{-\gamma_2 \tau})^{\frac{1}{2}} \times [a-b e^{-{\frac{1}{2}}(\gamma_1+\gamma_2)\tau}]$
 \cite{thila}.

The  density matrix in  Eq.(\ref{matrix1}) yields
$ S(\rho_{_{\mathrm{qi}}}) = -b^2 u_i^2 \log_2[b^2 u_i^2]-
(a^2+b^2 v_i^2) \log_2[a^2+b^2 v_i^2] (i=1,2)$ with  explicit dependence
on  the measurement time duration $\tau$, and system bias, 
 $\Delta \Omega$ and tunneling  amplitude
$\Delta$  via the functions $u_i$.
The condition $ S(\rho_{_{\mathrm{q1}}})=S(\rho_{_{\mathrm{q2}}})$
is therefore satisfied only if these parameters are the same for both
subsystems.
The quantum discord present in the two-qubit  
($\mathcal{D}_{_{\mathrm{q1,q2}}}(\tau)$) and 
two-reservoir ($\mathcal{D}_{_{\mathrm{r1,r2}}}(\tau)$) partitions for the subclass of
density matrix for which $\gamma_1 = \gamma_2$ (i.e. $f_2=f_3$) are
evaluated following  Fancini et al. \cite{fan}. We obtain
$\mathcal{D}_{_{\mathrm{q1,q2}}}(\tau) = H(b^2 u^2)-  H(\frac{1}{2}(1+ (1-4 b^2 u^2 v^2)^{1/2})$,
 and from which $\mathcal{D}_{_{\mathrm{r1,r2}}}(\tau)$ is obtained   by swapping 
 $u \leftrightarrow v$. The function  $H(x)=-x \log_2x-(1-x) \log_2(1-x)$,
 and the difference in quantum discords,
$\mathcal{D}_{_{\mathrm{q1,q2}}}-\mathcal{D}_{_{\mathrm{r1,r2}}}
= H(b^2 u^2)-H(b^2 v^2)$. 

For $\gamma_1 \neq \gamma_2=\gamma$ or unequal $u_1, u_2$ values, the quantum discord of the 
density matrix in  Eq.(\ref{matrix1}) can  evaluated  following
the main results in  Ref. \cite{ali,ali2} where the 
quantum conditional entropy is generalized as
$S(\rho|\{\Pi_k\}) = p_0 \, S(\rho_0) + p_1 \, S(\rho_1)$,
based on the earlier work of Luo \cite{luo}.
The terms  $p_0 = [(f_1 + f_3)k + (f_2 + f_4)l]$,
$p_1 =[ (f_1+ f_3)l + (f_2 + f_4)k]$ and 
$S(\rho_0), \, S(\rho_1)$ are dependent on 
generalized angles $\theta, \theta'$. 
 The generally cumbersome  procedure of determining 
$S(\rho|\{\Pi_k\})$ and the classical correlation
is  greatly simplified if  cross terms $\rho_{23}=0$ (following the notation 
in Ref.\cite{ali}) as is the case in the
density matrix in  Eq.(\ref{matrix1}). The
 problem reduces to minimization with  just one parameter $k$ or $l=1-k$,
instead of the set of three parameters.

\begin{figure}[htp]
  \begin{center}
    \subfigure{\label{fig1c}\includegraphics[width=4.15cm]{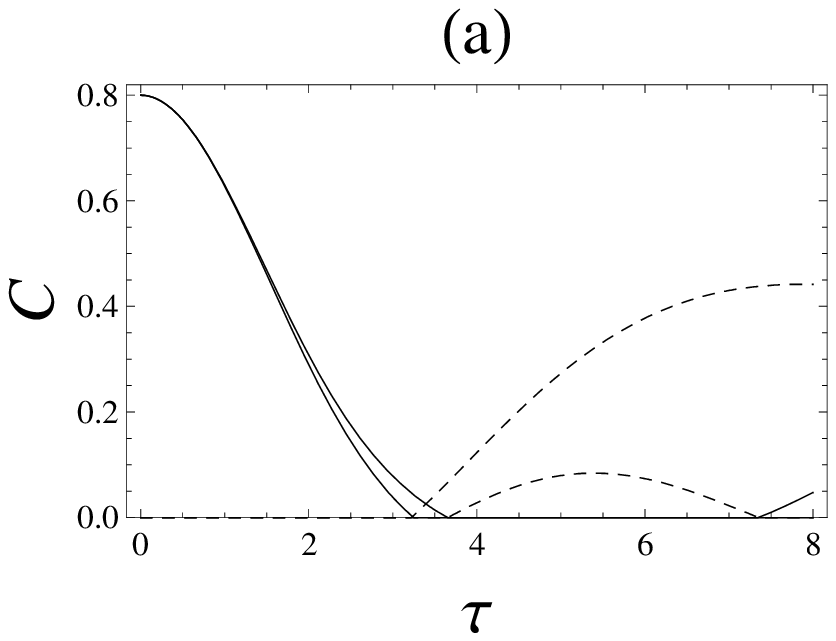}}\vspace{-1.1mm} \hspace{1.1mm}
     \subfigure{\label{fig1d}\includegraphics[width=4.15cm]{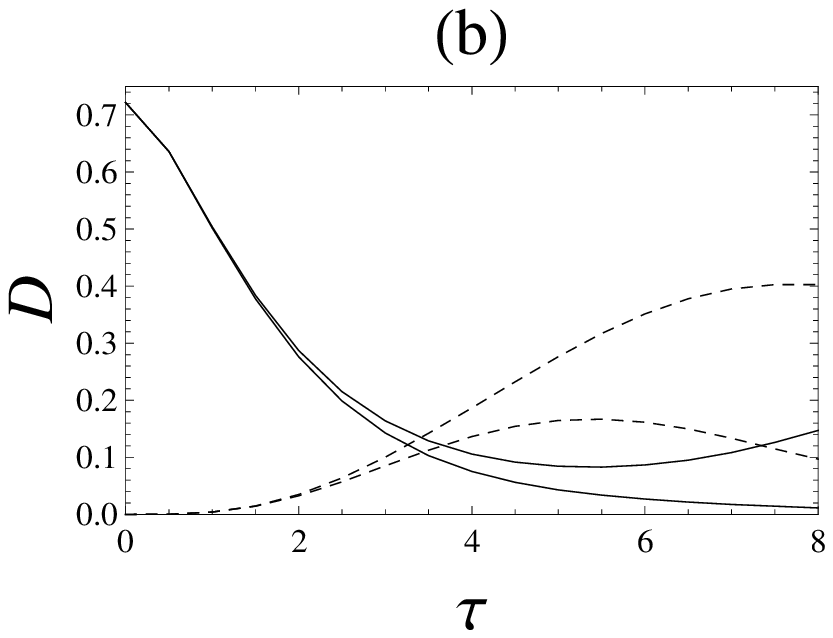}}\vspace{-1.1mm} \hspace{1.1mm}
     \end{center}
  \caption{a)  Two-qubit concurrence
$\mathcal{C}_{_{\mathrm{q1,q2}}}(\tau)$ (solid line)
as function of measurement time duration  $\tau$  with  initial amplitude parameter 
$a=\sqrt{1/5}$ with  same subsystem bias
$\Delta \Omega_1$=$\Delta \Omega_2$=$0.65$ (upper solid line),
dissimilar  subsystem bias
$\Delta \Omega_1$=$0.65, \Delta \Omega_2$=$0.15$ (lower solid line),
tunneling amplitude $\Delta$=$0.6$, $\eta$=$0.05$, $\omega_c$=$1$. 
The two-reservoir concurrence $\mathcal{C}_{_{\mathrm{r1,r2}}}(\tau)$ is
computed for  
$\Delta \Omega_1$=$\Delta \Omega_2$=$0.65$ (lower dashed line) 
and $\Delta \Omega_1$=$0.65, \Delta \Omega_2$=$0.15$ (upper dashed line).
 \\
b) Quantum discord  present in the two-qubit partition, 
 $\mathcal{D}_{_{\mathrm{q1,q2}}}(\tau)$  (solid lines)
and two-reservoir discord $\mathcal{D}_{_{\mathrm{r1,r2}}}(\tau)$ (dashed lines)
as function of measurement time duration  $\tau$. All other parameters and positioning
of lines with respect to bias configurations 
are the same as in (a).  \\
}
 \label{fig1}
\end{figure}

Figures~\ref{fig1}a,b show the notable differences, in the context of the
Zeno effect, between the  Wootters concurrence and the quantum discord.
For similar bias configuration,  $\Delta \Omega_1$=$\Delta \Omega_2$=$0.65$, 
and at the  initial state  parameter $a=\sqrt{1/5}$ (see Eqs.(\ref{fstate2})),
 the   qubit-qubit concurrence displays death and  rebirth events with increasing $\tau$, 
while the  reservoir-reservoir  concurrence is  short-lived. 
There is some departure from this trend for the 
dissimilar bias configuration,  $\Delta \Omega_1$=$0.65, \Delta \Omega_2$=$0.15$,
with  no rebirth in qubit-qubit concurrence, and the 
reservoir-reservoir  concurrence persists for longer times.
This behavior  is in  stark contrast to the more resilient  quantum discord 
which clearly displays the transition point at the similar bias configuration 
in Figure~\ref{fig1}b. Due to coupling with a system of lower bias, 
a transition point is 
not present at the dissimilar bias configuration. 

The crossover or transition point which occurs at the
 minimum (maximum) in the  two-qubit partition 
(two-reservoir partition) can be numerically verified 
using  Eqs.(\ref{eq:rateT}). We noted that 
the crossover point at a  Zeno/anti-Zeno transition
coincides with the equivalent point for the quantum 
discord, thus a decrease to increase and then a 
subsequent decrease in quantum discord can be interpreted 
as a sign of the Zeno/anti-Zeno transition.  

\begin{figure}[htp]
  \begin{center}
    \subfigure{\label{fig1c}\includegraphics[width=4.15cm]{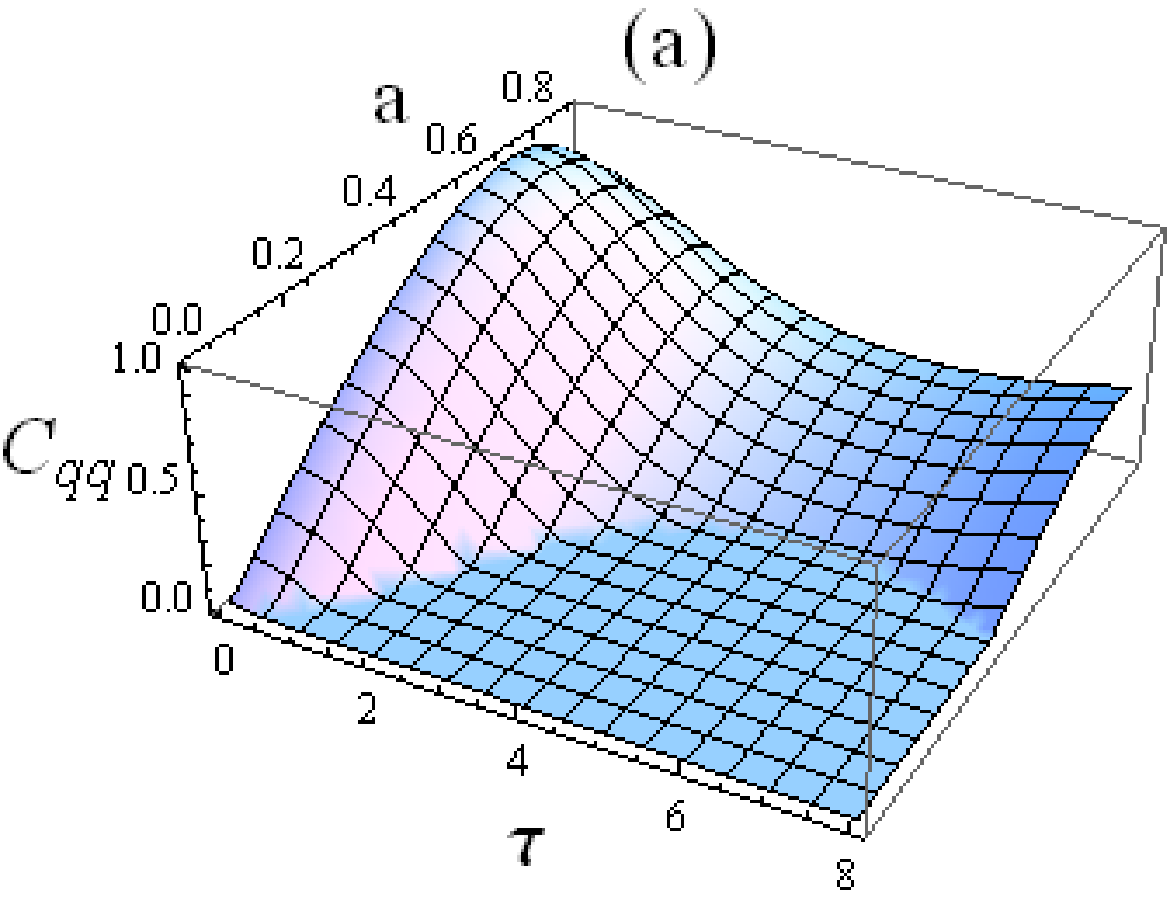}}\vspace{-2.1mm} \hspace{1.1mm}
     \subfigure{\label{fig1d}\includegraphics[width=4.15cm]{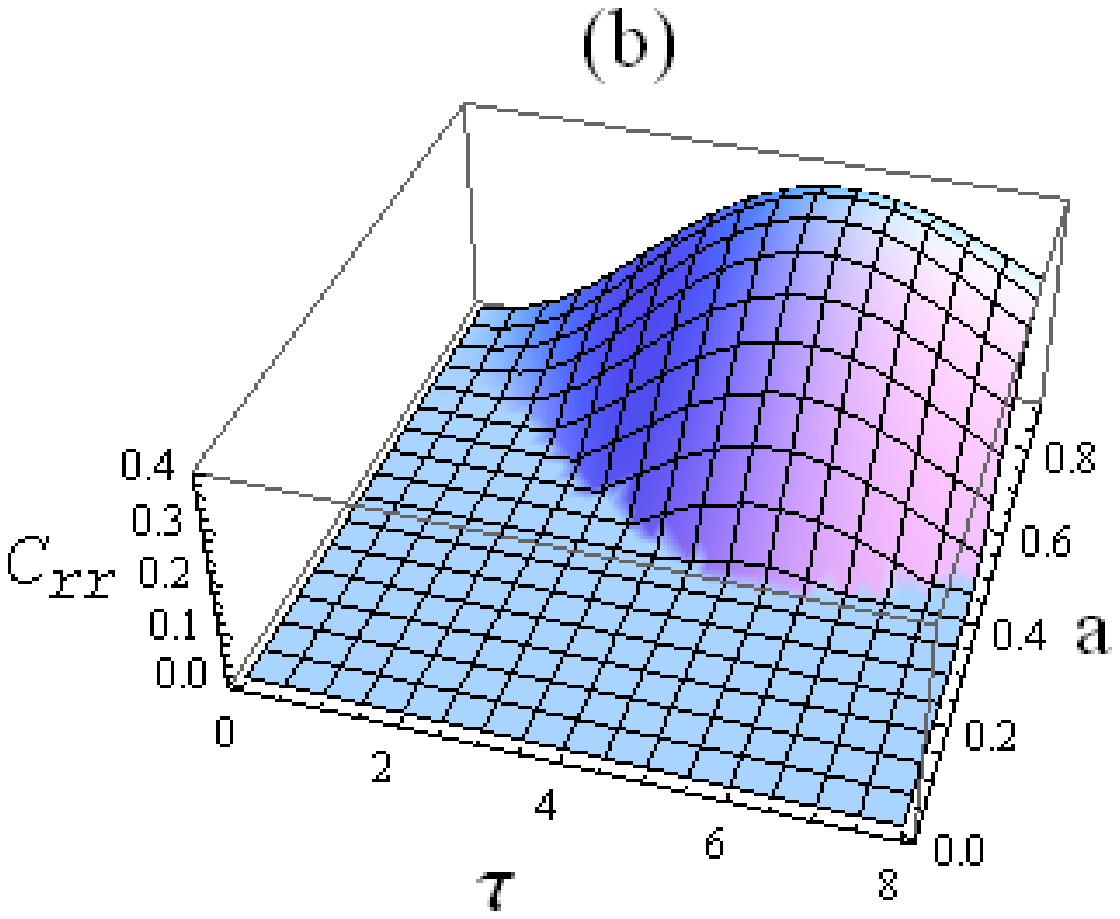}}\vspace{-2.1mm} \hspace{1.1mm}
 \subfigure{\label{fig1c2}\includegraphics[width=4.15cm]{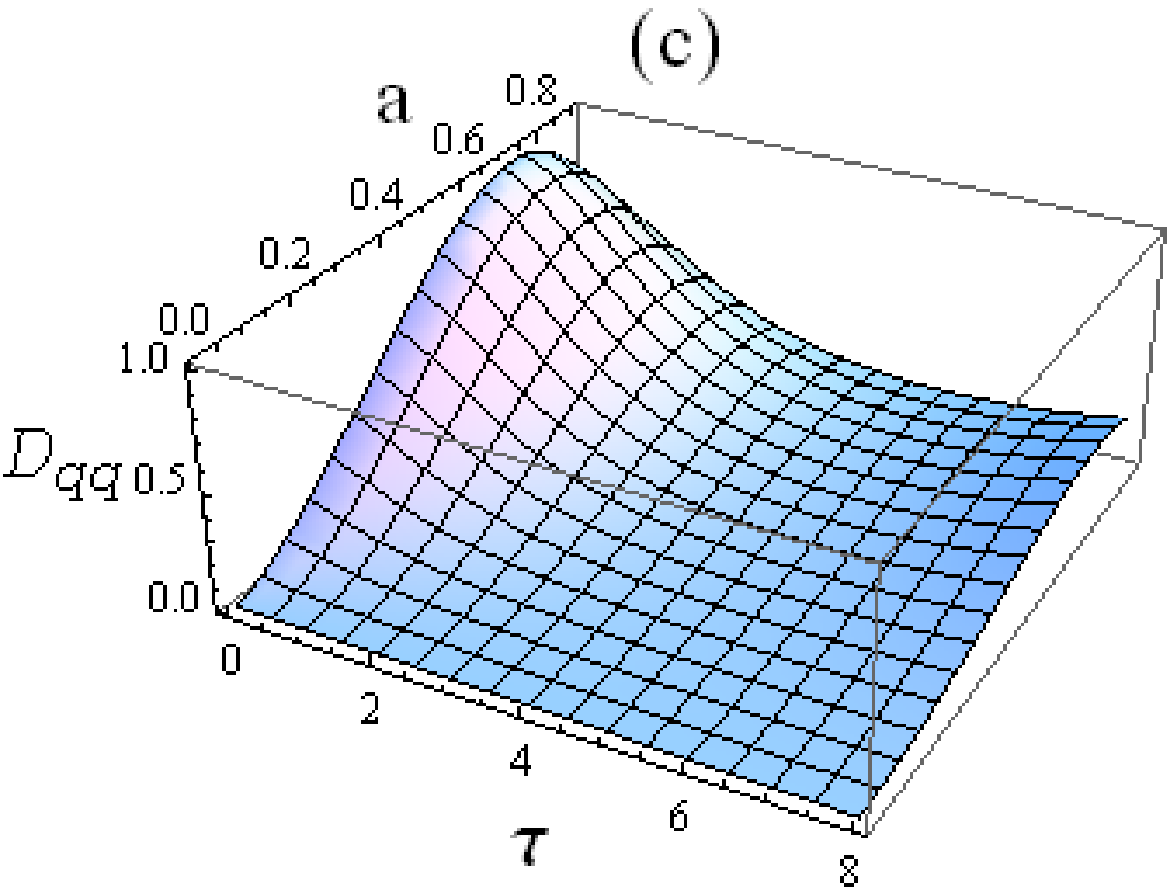}}\vspace{-2.1mm} \hspace{1.1mm}
     \subfigure{\label{fig1d2}\includegraphics[width=4.15cm]{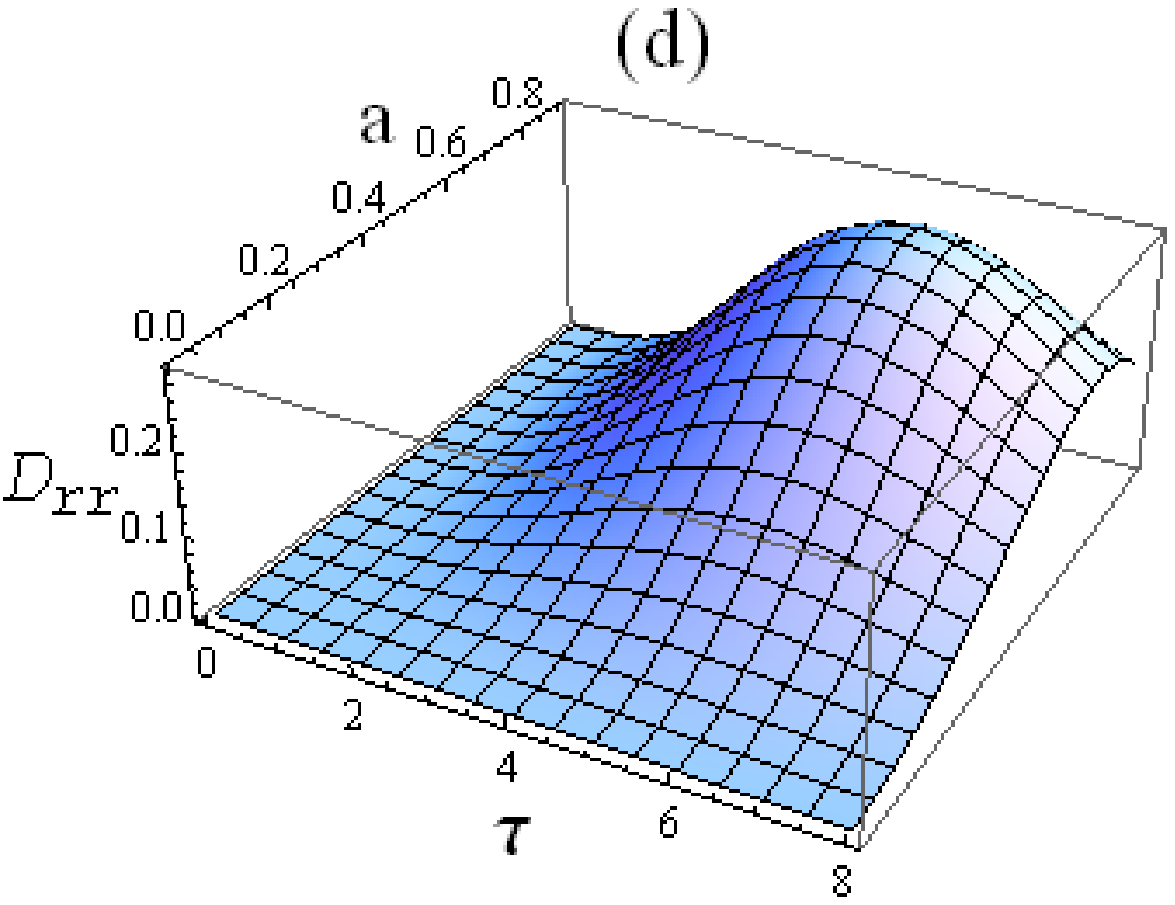}}\vspace{-2.1mm} \hspace{1.1mm}
     \end{center}
  \caption{(a),(b) Two-qubit concurrence
$\mathcal{C}_{_{\mathrm{q1,q2}}}(\tau)$  and
two-reservoir concurrence $\mathcal{C}_{_{\mathrm{r1,r2}}}(\tau)$
as function of measurement time duration  $\tau$  and   initial amplitude parameter 
$a$ with  same subsystem bias
$\Delta \Omega_1$=$\Delta \Omega_2$=$0.65$. All other parameters
are the same as in Figures~\ref{fig1}a,b. \\ (c),(d)
Quantum discord  present in the two-qubit partition, 
 $\mathcal{D}_{_{\mathrm{q1,q2}}}(\tau)$  
and two-reservoir discord $\mathcal{D}_{_{\mathrm{r1,r2}}}(\tau)$ 
as function of measurement time duration  $\tau$. All other parameters 
are the same as in Figures~\ref{fig1}a,b.  
}
 \label{fig2}
\end{figure}

Figures~\ref{fig2}a,b,c,d which incorporates 
a change in the  initial state  parameter $a$,
show that the two-reservoir discord best captures the 
Zeno- anti Zeno transition point. 
While it is known that the quantum discord remains non-zero under
various conditions \cite{zu,ve1,ve2}, these results 
show that the quantum discord is  
reliable in being  able to display Zeno-anti-Zeno 
dynamics occurring in separate qubit-reservoir subsystems, and  which  are 
also weakly coupled (small values of $a$).

\subsection{\label{dyn2nd} 
Quantum discord in an initial state with single excitation}

The analysis of quantum discord can be extended to 
the initial  state of the Bell-like state with 
just a single excited state residing in either of the qubit
\begin{eqnarray}
\ket{\Phi}_0 &=& \left[ c\ket{0}_{\mathrm{ex1}}\ket{1}_{\mathrm{ex2}}
+ d\ket{1}_{\mathrm{ex1}}\ket{0}_{\mathrm{ex2}} \right]
 \ket{0}_{\mathrm{r1}} \ket{0}_{\mathrm{r2}},  
\label{gstate2}
\end{eqnarray}
 where $i$=$1,2$ denote the two qubit-reservoir systems with associated
functions $u_i(t)$. As in the case in Eq.(\ref{matrix1}),
we   trace  out the reservoir states to  obtain
a  time-dependent qubit-qubit reduced density matrix 
\begin{equation}\label{matrix2}
\rho_{_{\mathrm{q1,q2}}}(t)=\left(
\begin{array}{cccc}
 g_1 & 0 & 0 & 0 \\
  0 & g_2 & g_4 & 0 \\
  0 & g_4 & g_3 & 0 \\
  0 & 0 & 0 & 0\\
\end{array}
\right).
\end{equation}
where for $t \ge 0$, the matrix elements evolve as 
$g_1(t)= c^2 v_2(t)^2 + d^2 v_1(t)^2$,
$g_2(t)= c^2 u_2(t)^2$, $g_3(t)= d^2 u_1(t)^2$ and
$ g_4(t)=c d u_1(t) u_2(t)$. 
Following  Fancini et al. \cite{fan}, we obtain
$\mathcal{D}_{_{\mathrm{q1,q2}}}(\tau) = H(a^2 u^2)-  H(u^2) + 
H(\frac{1}{2}(1+ (1-4 b^2 u^2 v^2)^{1/2})$,
 from which $\mathcal{D}_{_{\mathrm{r1,r2}}}(\tau)$ is obtained   by swapping 
 $u \leftrightarrow v $ and  
 the difference in quantum discords, 
$\mathcal{D}_{_{\mathrm{q1,q2}}}-\mathcal{D}_{_{\mathrm{r1,r2}}}
= H(a^2 u^2)+H(v^2)-H(u^2)-H(a^2 v^2)$.

\begin{figure}[htp]
  \begin{center}
    \subfigure{\label{fig2a}\includegraphics[width=4.1cm]{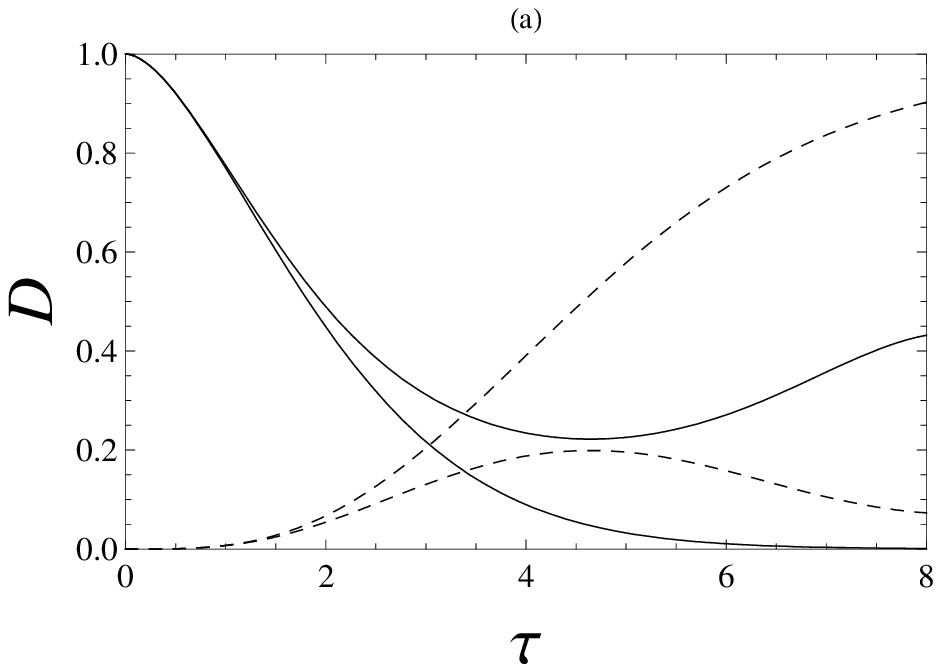}}\vspace{-3.1mm} \hspace{1.1mm}
     \subfigure{\label{fig2b}\includegraphics[width=4.1cm]{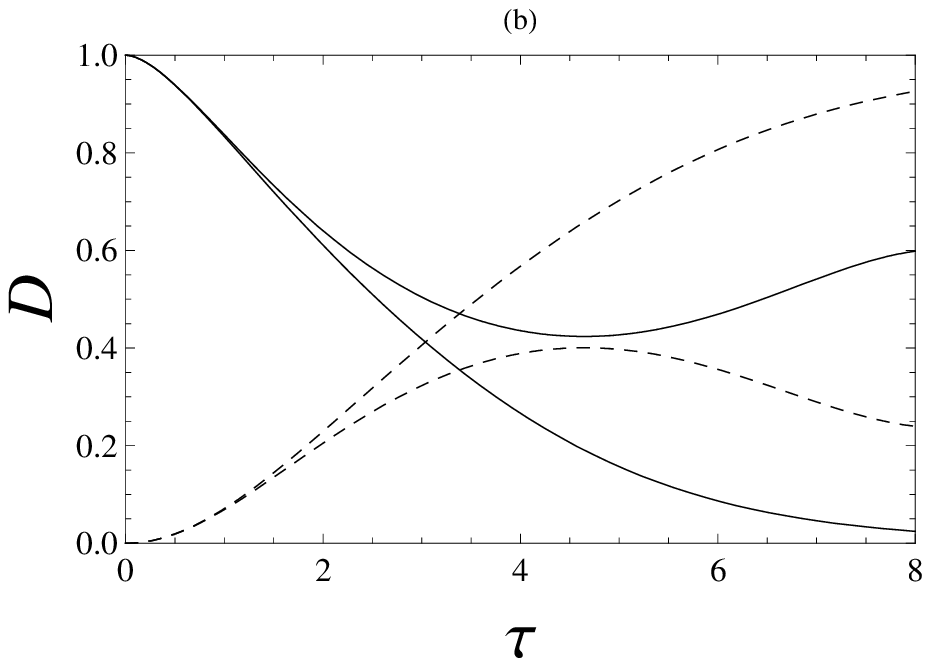}}\vspace{-3.1mm} \hspace{1.1mm}
     \end{center}
  \caption{a) Quantum discord  in the two-qubit partition,  
$\mathcal{D}_{_{\mathrm{q1,q2}}}(\tau)$  (solid lines) 
as function of measurement time duration   $\tau$  with  same subsystem bias
$\Delta \Omega_1$=$\Delta \Omega_2$=($0.75$, upper solid line and 
$0.25$, solid line), tunneling amplitude 
$\Delta$=$0.6$, $\eta$=$0.05$, $\omega_c$=$1$.
Quantum discord present in the two-reservoir partition,
 $\mathcal{D}_{_{\mathrm{r1,r2}}}(\tau)$  is denoted 
 by dashed lines, $\Delta \Omega_1$=$\Delta \Omega_2$=$0.75$ corresponds to the lower dashed line,
$\Delta \Omega_1=\Delta \Omega_2$=$0.15$ corresponds to the 
 upper dashed line. All measures are evaluated using the initial state in 
Eq.(\ref{fstate2}) with initial amplitude parameter $a=\sqrt{1/2}$. 
\\
b) Quantum discord  present in the two-qubit partition, 
 $\mathcal{D}_{_{\mathrm{q1,q2}}}(\tau)$  (solid lines)
and two-reservoir discord $\mathcal{Q}_{_{\mathrm{r1,r2}}}(\tau)$ (dashed lines)
as function of measurement time duration  $\tau$. All other parameters and positioning
of lines with respect to bias configurations 
are the same as in (a). All measures are evaluated using the initial Bell-like state in 
Eq.(\ref{gstate2}) with initial amplitude parameter $c=\sqrt{1/2}$. 
}
 \label{fig23}
\end{figure}

Figures~\ref{fig23}a,b show  the dynamics of the two-qubit quantum discord,  
$\mathcal{D}_{_{\mathrm{q1,q2}}}(\tau)$  and  two-reservoir
quantum discord $\mathcal{D}_{_{\mathrm{r1,r2}}}(\tau)$ at  different
 subsystem bias configurations, 
$\Delta \Omega_1$=$\Delta \Omega_2$=$0.75$, $0.25$.
The quantum discord displays 
anti-crossing  behavior at the higher system bias value 
for the two different states given in Eqs.(\ref{fstate2}) 
and (\ref{gstate2}). The  
 two-reservoir quantum discord is however enhanced  in  
Eq. (\ref{gstate2}), due to greater participation from 
the two-reservoir partition. The slight differences in the quantum
discord due to the two different initial states in  
Eqs.(\ref{fstate2}), (\ref{gstate2}) are mainly due to 
variations in classical correlations, $\mathcal{C}_i(\rho)$ 
where $i$ denotes the subsystem under consideration.

\section{Quantum discord and exceptional points at 
 high precision measurements}\label{exp}

While the quantum Zeno effect is  viewed
as the effect of repeated measurements  on a quantum
system, it can be studied in the wider  context of 
the dynamical time evolution of quantum systems.
The Zeno effect appears even if  the information regarding 
the state of the observed system manifests in the form
of an  external degree of freedom such as the 
 spontaneous emission process. 
It would be interesting to examine whether the features of the Zeno
effect, and the quantum discord  are retained if the
 monitoring device imparts
a significant  disturbance on the system under study and itself  
dominates the time evolution of the quantum
system.

For a two-level system with energies $E_1$ ($E_2$) at state $\ket{0}$
($\ket{1}$) subjected to a continuous measurement process,   its original 
Hamiltonian  $\widehat  H_0$ transforms via the non-Hermitian Hamiltonian
\cite{men,ono} $\widehat H_{eff} = \widehat H_0-
i{\hbar\over{\tau E_r^2}}(\widehat H_0-E)^2$.
$E$  is the selected measurement output  after a time $\tau$ and 
$E_r$ is the error made during the measurement of the energy, $E$.
$E_r$ can also be considered as a measure of the precision of the
monitoring device.
A large error made during the measurement can be viewed 
as a weak or unsharp measurement and $\widehat H_{eff} \rightarrow 
\widehat  H_0$, whereas one made with very small error can be
considered a highly precise measurement.
The system therefore evolves
as $i \hbar {\partial \over \partial t}\ket{\psi(t)}= 
H_{eff} \ket{\psi(t)}$ during measurement due to the constraining effect of the 
 selected readout $E$. 

The state of the  system being measured can be expanded 
 within the unperturbed basis states $\ket{n}$
of the unmeasured system with Hamiltonian $\widehat H_0$ as
$|\psi(t) \rangle =\sum_n C_n(t) |n \rangle$. The coefficients
$ C_n(t)$ can be determined \cite{ono,gar,staf} using the Schr\"odinger equation
and  the non-Hermitian $\widehat H_{eff}$.
In the presence of an external potential 
of the form $V_{22}=V_{11}=0$ and
$V_{12} = V_{21}^\ast = V_0 e^{i \omega t}$ with $V_0$ real,
the system evolves as $\ket{\psi(t)} = e^{-i(E_1-2 i \lambda_1) t} C_1(0)\ket{0}
+ e^{-i(E_2-2 i \lambda_2) t} C_2(0)\ket{1}$
where $\lambda_1$=$\frac{(E_1-E)^2}{2 \tau E_r^2}$ and
$\lambda_2$=$\frac{(E_2-E)^2}{2 \tau E_r^2}$.

The coefficients  $C_1(t), C_2(t)$ can be recast as
\begin{equation}
 \left[ \begin{array}{c}
  C_1(t) \\
  C_2(t) \\   \end{array} \right] =
\left[
  \begin{array}{cc}
    \cos {\kappa}t-i \alpha_1 
&-i\alpha_2\\
    -i\alpha_2 & \cos {\kappa} t+i\alpha_1 
\\  \end{array}  \right]\
\left[ \begin{array}{c}
    C_1(0) \\   C_2(0) \\
  \end{array}
\right],
\end{equation}
where $\alpha_1$=$\cos \theta \sin{\kappa} t$, 
 $\alpha_2$=$\sin \theta \sin{\kappa} t$,
  $\cos \theta$=$\frac{q}{\kappa}$, 
$\kappa$=$\sqrt{q^2+p^2}$, $q$=$\frac{1}{2}(\omega-\Delta E+2 i \Omega)$, 
$\Delta E$=$(E_2-E_1)$,  $p$=$V_0$ and  $\Omega$=$\lambda_2$-$\lambda_1$. The  terms 
$\lambda_2$ and $\lambda_1$ as defined in the earlier paragraph
are dependent on the measurement precision, $E_r$ as well as the
energy $E$ to be measured. 

For a system in which the initial state at $t=0$ is 
 $\ket{1}$ and the final state 
at time $t$ is either $\ket{1}$ or $\ket{0}$,
the probability $P_{11}$ of the 
system to be in the state  $\ket{1}$ 
can be obtained following Ref.\cite{staf} as 
$P_{11} = 
 |\cos^2{\kappa} t-i \cos\theta \sin{\kappa t}|^2 e^{-\lambda_t t}$
 where $\lambda_t$=$\frac{(E_2-E_1)^2}{2 \tau E_r^2}$.
Likewise the probability $(P_{10}) $ that the system is present in 
the state  $\ket{0}$ is given by $P_{10} 
= |\sin^2\theta\,\sin^2{\kappa t}|^2 e^{-\lambda_t t}$.
The total probabilities, $P_{11}$+$P_{10} \le 1$,
the loss of normalization is dependent on the measurement
precision, $E_r$ as expected, and further evaluation of
the quantum discord will be significantly affected in the case of
highly precise measurements.

At  the resonance frequencies, $\omega =\Delta E$, 
the Rabi frequency $\kappa_0=2 (V_0^2 -\lambda_t^2)^{1/2}$, and 
$\cos\theta = -i\lambda_t/\kappa_0$. There are two 
 regimes, depending on the relation between $V_0$ and $\lambda_t$. The range where 
$V_0 > \lambda_t$ applies to the coherent tunneling regime where
\begin{eqnarray}
\nonumber
P_{11} &=&  e^{-\lambda_t t}
\left[\cos{\kappa_0} t- \frac{\lambda_t}{\kappa_0}\sin{\kappa_0} t \right]^2 
\\ \label{eq:co}
 P_{10} &=&   e^{-\lambda_t t} \frac{V_0^2}{\kappa_0^2}\sin^2{\kappa_0} t,
\end{eqnarray}
For $V_0< \lambda_t$, the system undergoes incoherent tunneling
\begin{eqnarray}
\nonumber
P_{11} &=&  e^{-\lambda_t t}\left[\cosh{\kappa_0} t- 
\frac{\lambda_t}{\kappa_0}\sinh{\kappa_0} t\right]^2,
 \\ \label{eq:inco}
P_{10} &=&   e^{-\lambda_t t} \frac{V_0^2}{\kappa_0^2}\sinh^2{\kappa_0} t
\end{eqnarray}
At the exceptional point, $\kappa_0 = 0$, and both regimes merge and we obtain
$P_{11} =  \left(1-  \frac{\lambda_t t}{2}\right)^2 e^{-\lambda_t t}$ and
$ P_{10} = \left(\frac{\lambda_t t}{2}\right)^2 e^{-\lambda_t t}$.
Exceptional points are  singularities \cite{Heiss} which appear at the branch point
of eigenfunctions due to changes in parameters which govern the 
non-Hermitian Hamiltonian operator. These 
points are known to be  located  in the vicinity
of a level repulsion \cite{Heiss} and unlike degenerate points,
only one eigenfunction exists at the exceptional point due to the merging of
two eigenvalues. In the case of the quantum measurements considered in this
work, the exceptional point 
appears at a critical measurement precision $E_r^c$=$\frac{\Delta E}{\sqrt{2 \tau V_0}}$.
Considering a unit system in which $\hbar=V_0=\Delta E$=1, $\tau=2 \pi/V_0$ and a unitless 
time $t=t'/\tau$,  we obtain  $E_r^c=\frac{1}{\sqrt{4 \pi}}$. Using
$r$ to denote the unitless measurement precision parameter, we note that
at  $r > \frac{1}{\sqrt{4 \pi}}$ ($r < \frac{1}{\sqrt{4 \pi}}$), the quantum system undergoes coherent (incoherent) tunneling.

\subsection{ Entangled qubits   subjected 
to high precision measurements}\label{prec} 

Similar to the model adopted in the Section \ref{dyn}, 
we consider two  uncoupled qubits  which are entangled initially, but which 
differ from the earlier treatment in being monitored by independent 
observers. These observers  assume the role of the reservoirs  of harmonic oscillators.
We consider  functions $u(t)$ and $v(t)$  which previously
 were associated with the decay of the two-level qubit in Eq.(\ref{fstate}).
The influence of the measurement precision 
 $E_r$ on the quantum discord is investigated by 
setting $u(t)^2$ = $P_{11}$, $v(t)^2 = P_{10}$,
and evaluating $\mathcal{D}_{_{\mathrm{q1,q2}}}(t)$
and $\mathcal{D}_{_{\mathrm{r1,r2}}}(t)$ as described in Section \ref{dyn}.
Unlike  in Sections \ref{ca}, \ref{c1b}, here we examine the dynamics of 
the quantum discord in the context of a Zeno effect
manifesting itself even before a measurement outcome is reached
and therefore time, $t$ satisfies $0 \leq t \leq \tau$ where $\tau$ is the measurement
duration.

\begin{figure}[htp]
  \begin{center}
    \subfigure{\label{figzca}\includegraphics[width=4.1cm]{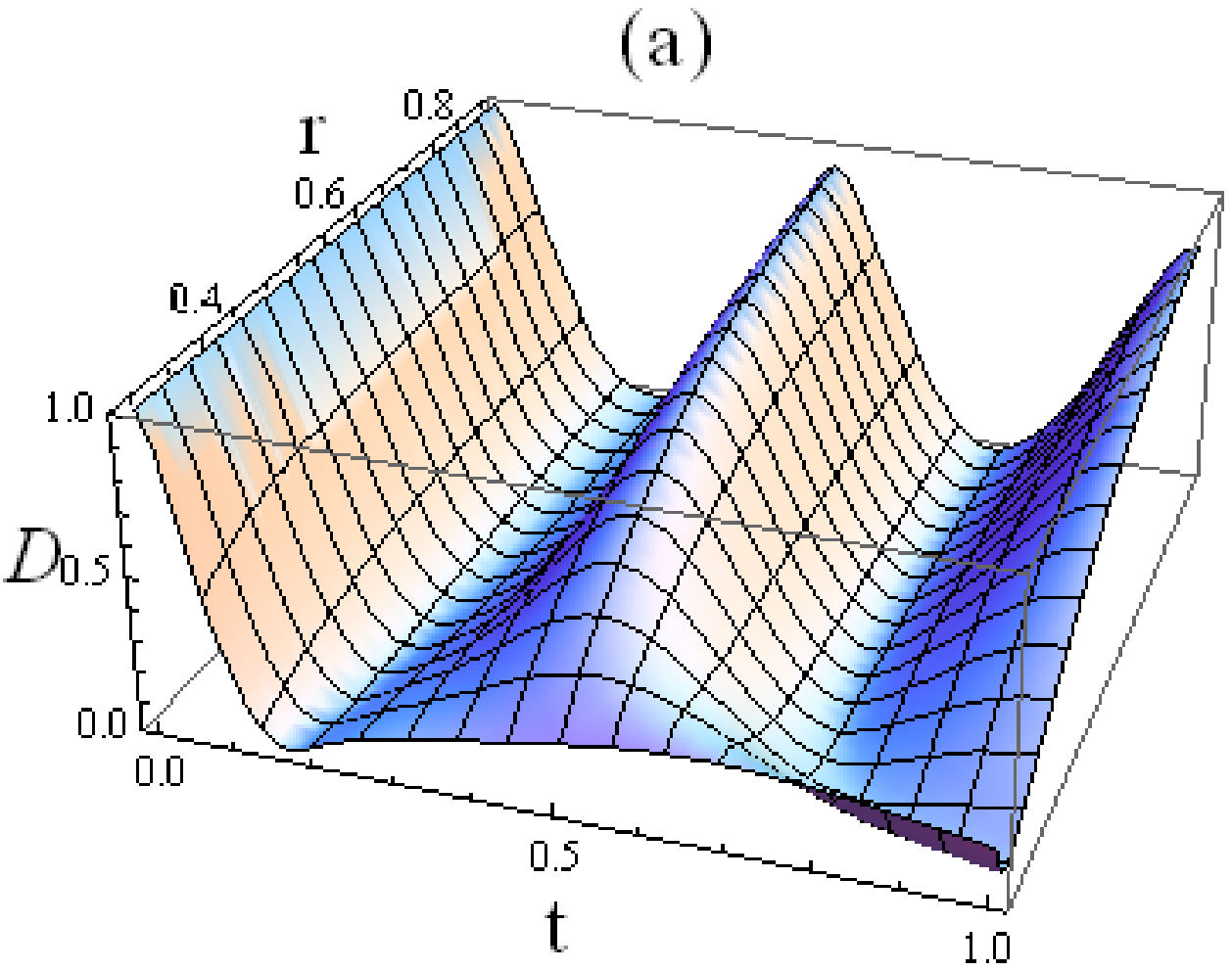}}\vspace{-3.1mm} \hspace{1.1mm}
     \subfigure{\label{figzcb}\includegraphics[width=4.1cm]{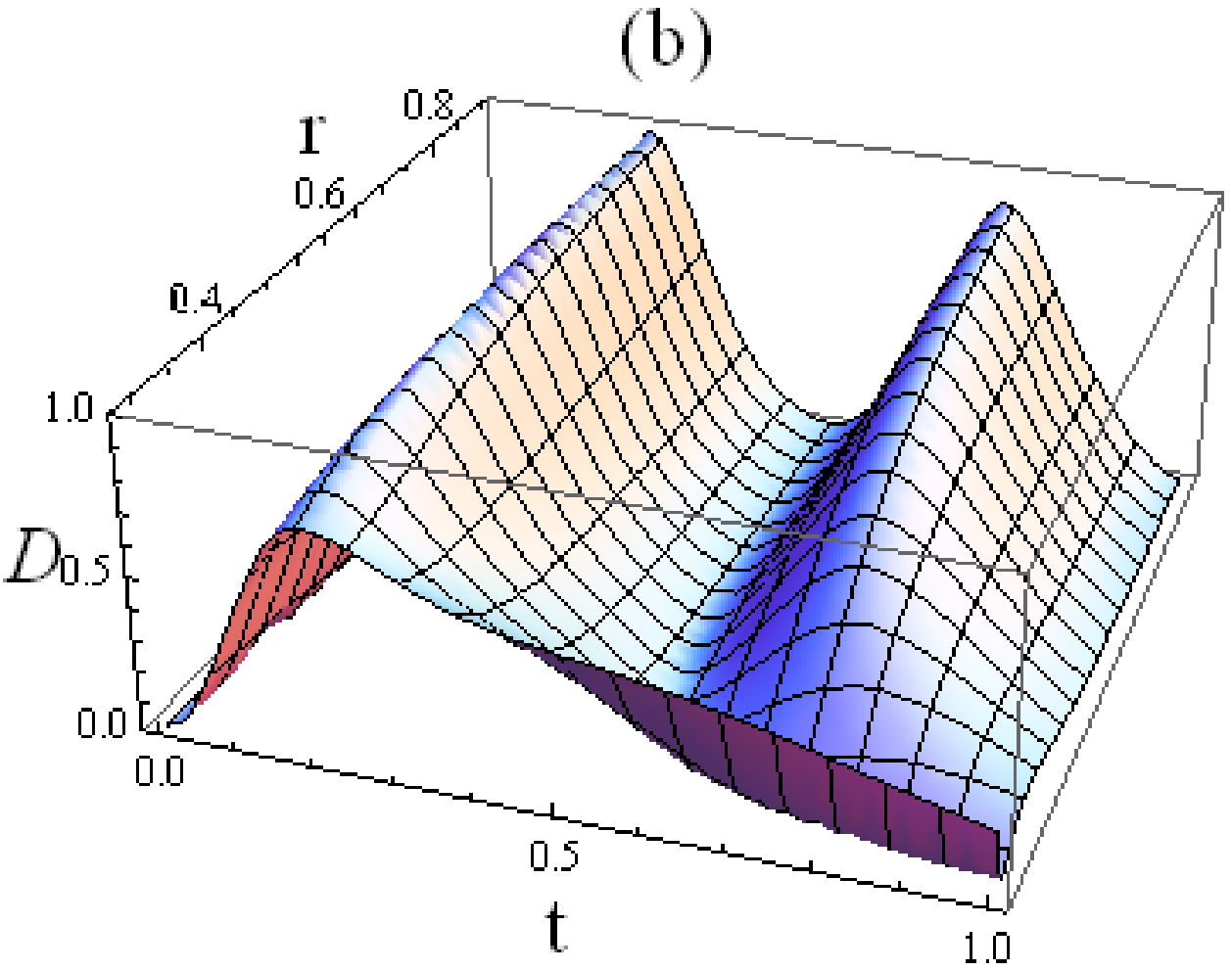}}\vspace{-3.1mm} \hspace{1.1mm}
     \end{center}
  \caption{a) Quantum discord $\mathcal{D}$ present in the two-qubit partition,  
 as function of normalized time  $t$  and 
measurement precision $r$ in the coherent tunneling regime. The  initial amplitude parameter 
$a$ is set at 0.7 in Eq.(\ref{fstate2}), 
with $\hbar=V_0=\Delta E$=1, $\tau=2 \pi$ and a unitless 
time $t=t'/\tau$, \\
b) Quantum discord $\mathcal{D}$ present in the two-reservoir partition,
 as function of normalized time   $t$  and
$r$.  All other parameters  
are the same as in (a).
}
 \label{figL}
\end{figure}

\begin{figure}[htp]
  \begin{center}
    \subfigure{\label{figzca}\includegraphics[width=4.1cm]{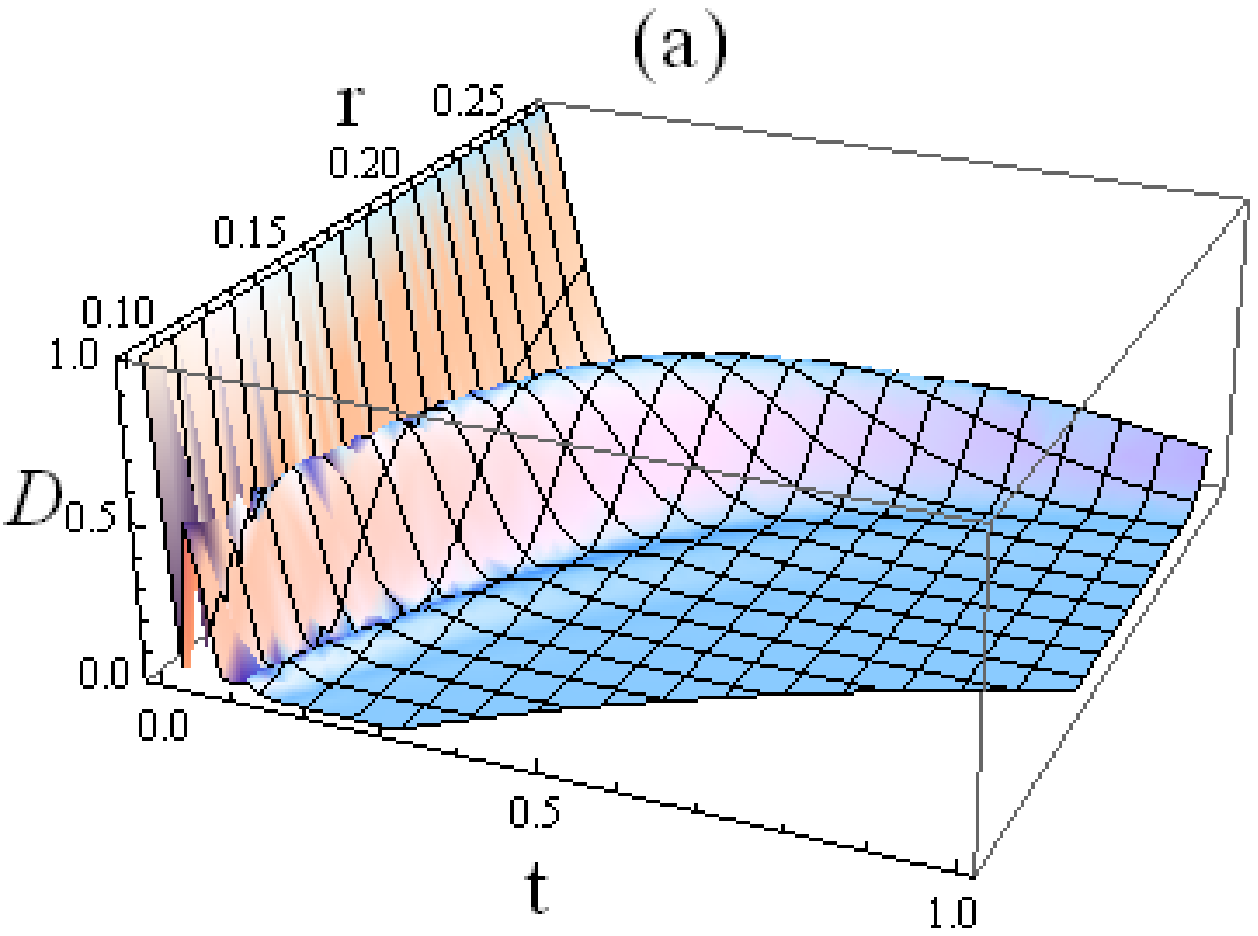}}\vspace{-3.1mm} \hspace{1.1mm}
     \subfigure{\label{figzcb}\includegraphics[width=4.1cm]{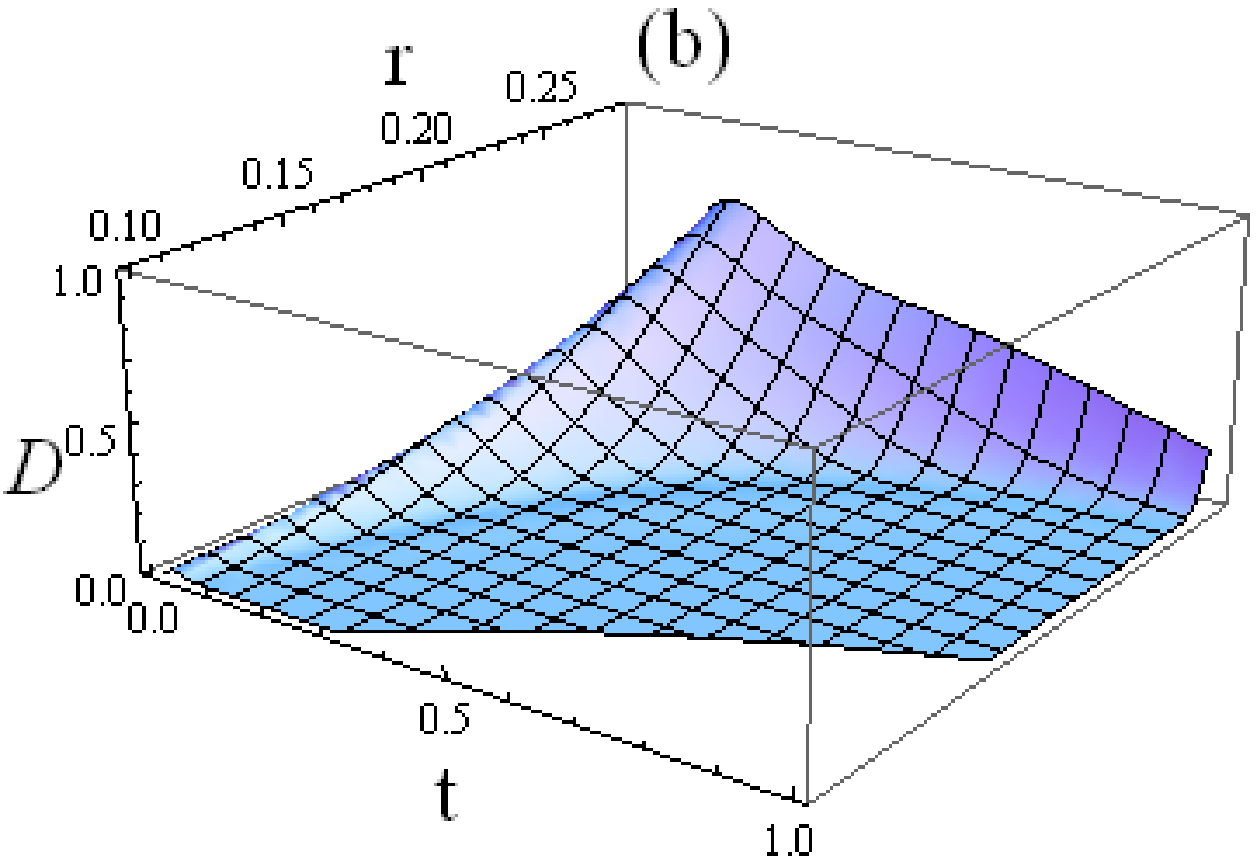}}\vspace{-3.1mm} \hspace{1.1mm}
     \end{center}
  \caption{a) Quantum discord $\mathcal{D}$ present in the two-qubit partition,  
 as function of normalized time   $t$  and measurement precision $r$ 
in the incoherent tunneling regime. All parameters  
are the same as in Figures~\ref{figL}a.\\
b) Quantum discord $\mathcal{D}$ present in the two-reservoir partition,
 as function of normalized time    $t$  and
measurement precision $r$.  All other parameters  
are the same as in (a). The white region 
 corresponds to indeterminate values of the quantum discord.}
 \label{figU}
\end{figure}

As the  relation $u(t)^2 + v(t)^2=1$ is not satisfied
for high precise measurements, the widely accepted
defination of the quantum discord discussed in Section \ref{meas}
may be considered as a limiting case of a more   generalized defination
that may apply in the case of quantum systems which undergo non-Hermitian evolution dynamics.
With the inclusion of a non-Hermitian term, the  unit trace
and strict positivity conditions of the density operator
of the quantum system will not be satisfied as well. 
With the view of realizing qualitative results of the quantum discord,
we therefore relax conditions needed for more rigorous and accurate
quantitative approach to evaluating the quantum discord for non-Hermitian systems.
  We illustrate the dynamics of the quantum discord in the two regimes
 specified by  Eqs.(\ref{eq:co}) and (\ref{eq:inco})  in 
Figures~\ref{figL}a,b  and \ref{figU}a,b. 
These figures show the explicit dependence of the quantum discord
on the measurement precision,  with appearance of indeterminate values 
of the quantum discord at very high precision measurements (low values of
$r$). The figures also indicate that a highly precise observer
can diminish the non-classical correlation shared between two subsystems,
with the tendency to do so increasing with the meaurement precision.
 It has to be noted that the quantum discord is evaluated
in a reference frame where the observer is not under active consideration
as one of the subsystems. The results will therefore be modified if
the monitoring system is included and the quantum system then expands
to a group of three subsystems.

\section{Discussion and Conclusions}\label{con}

We have  examined  the influence of quantum 
measurements on quantum discord with 
consideration of  two types of measurements, 
weak or low precision measurements and highly precise measurements.
In the case of ideal weak measurements, the results show that the 
 quantum discord present in a  two qubit or reservoir 
system responds to  characteristic parameters 
such as the system  bias, duration and frequency of the measurement induced by the decoherence
processes as well as the strength of initial entanglement between the 
two qubit systems. Unlike  the reservoir-reservoir 
 concurrence $\mathcal{C}_{_{\mathrm{r1,r2}}}(\tau)$,
its quantum discord counterpart is more resilient to 
changes in the measurement duration, $\tau$.
For weak measurements, the quantum discord therefore presents as 
a suitable measure to identify and quantify
Zeno-anti Zeno crossover dynamics in the spin-boson system.  
The quantum discord may be used as a reliable measure of quantum processes influenced by the
quantum Zeno effect such as quantum switching and  preparation of
decoherence-free states and cluster states \cite{nel}. 
Another potential application is the possibility of using  quantum discord as an efficiency 
 measure of the  purification of qubit states which occurs via extraction
of a pure state through   a series of  Zeno-like measurements \cite{puri}.
The model used in this work is generic to most quantum systems
which undergo Zeno-anti-Zeno crossover dynamics, and can therefore
be extended to other quantum systems \cite{nosci,wangS,Segal,japko,rai}
 displaying such crossover effects, as mentioned earlier in the text.

For the class of highly precise measurements which introduce maximal 
interference in the dynamics of quantum systems, the appearance of
singularities introduce complications in the quantum evolution
of a measured system. The quantum discord
becomes indeterminate for highly precise quantum measurements. 
Importantly, the Zeno effect fails at very precise measurements as  the system does not reside at one  level, but possibly transfers available information to the unspecified 
level of the observer.
In future works, the direct influence of the 
Zeno effect due to measurements made in one subsystem
 in order  to  obtain the conditional entropy in a
 second subsystem will be considered. 
 Such an approach will allow determination of  the influence 
of the measurement precision on the classical correlation measure
in a neighboring partition. This alternative perspective of the 
influence of a monitoring device will also allow convenient analysis
of the Berry phase due to quantum measurements.

Finally, we have presented results of the influence 
of the quantum Zeno effect on the concurrence and
quantum discord for various biased configurations of the qubit-reservoir system.
We have  demonstrated  the resilience
of the quantum discord measure, in particular it is more robust than
the concurrence  in the reservoir-reservoir partition subsystem.
The quantum discord which is an intrinsically 
distinct entity from entanglement, therefore serves as a better indicator,
of the crossover point in 
Zeno to anti-Zeno transition  evident in some spin-boson systems
 under suitable conditions and for weak measurements. As to whether this applies to 
other quantum systems which display both Zeno and
anti-Zeno effects needs further investigation.  
For highly precise  measurements, 
the monitoring device can significantly interfere with the 
evaluation of  the  quantum discord and produce indeterminate values of
the quantum discord.   With progress in experimental
techniques and studies of quantum measurement in optics and nanostructure systems 
\cite{hall, tit},   investigations  involving the  quantum discord
of entangled systems are expected to play a greater role in future experimental 
works.

\section{Acknowledgments}
I am grateful to Prof. R. Onofrio for alerting me to Refs. \cite{men,ono}. I also 
acknowledge access to the National Computational Infrastructure 
 (NCI)   facilities which is supported by the Australian Commonwealth 
Government.

\end{document}